\documentstyle[aps,prd]{revtex}
\begin{document}
\draft
\title{Higher order terms in the contraction of SO(3,2)} 
\author{W. Smilga}
\address{Isardamm 135 d, D-82538 Geretsried, Germany}
\address{e-mail: wsmilga@compuserve.com}
\maketitle
\begin{abstract}
The contraction of a spin-1/2 representation of the de~Sitter group SO(3,2) 
yields a translation operator that consists of the usual momentum operator 
plus a second order term, the ``momentum spin" as described by F. G\"ursey. 
The contribution of momentum spin to the kinematics of a multiparticle system 
in a tangential space of anti de~Sitter space is analyzed. It is shown that
it can be described by a perturbation term with the structure of the 
interaction term of quantum electrodynamics. 
An evaluation of the corresponding coupling constant reproduces Wyler's
heuristic formula for the electromagnetic coupling constant.
\end{abstract}


\section{INTRODUCTION} 

The de~Sitter group SO(3,2) has been used for a long time as the group of 
basic symmetry transformations within a model of the universe that is defined 
by the anti de~Sitter space. 
The Poincar\'{e} group is then ``in some sense, a limiting case" of the 
de~Sitter group within a neighbourhood $\cal{N}$ of a given point $P$ of 
anti de~Sitter space if the ``de~Sitter radius" $R$ approaches infinity.
This limiting case has been mathematically formulated by E.~In\"on\"u and 
E.~P.~Wigner \cite{iw} in 1953 by the method of group contraction. 

This paper takes up a result that has been obtained three decades ago by 
several authors \cite{fg},\cite{ff} who studied the contraction of the 
de~Sitter SO(3,2) group and applied it to a multiparticle system in a 
tangential space to the de~Sitter space. 
  
Let $l_{ab}$, $a,b = 0,1,2,3,4$, be representations of the infinitesimal 
generators of SO(3,2) by operators in a quantum mechanical state space  
and let $l_{\mu4}$ be those operators that in the contraction limit converge 
towards the momentum operators $p_\mu$ of the Poincar\'{e} group. 
Then $l_{\mu\nu}$ are the operators of the Lorentz subgroup with 
$\mu,\nu = 0,1,2,3$. 
Group contraction assumes that the expectation values of $l_{\mu4}$ become 
very large so that for the amplitudes between states $\phi$ and $\phi'$
the following relation is valid
\begin{equation}
\langle\phi|l_{\mu4}|\phi'\rangle \;\; \gg \;\; 
\langle\phi|l_{\nu\rho}|\phi'\rangle . \label{1-1}
\end{equation}
As a consequence of the commutation relations of SO(3,2), the operators 
$l_{\mu4}$ can then be approximated by the translation operators $p_\mu$ of 
the Poincar\'e group $P(3,1)$. 

E.~In\"on\"u and E.~P.~Wigner (\cite{iw}, see also F. G\"ursey \cite {fg}) 
have performed the contraction process by rescaling the operators $l_{\mu4}$ 
\begin{equation}
\Pi_\mu = \frac{1}{R}\;l_{\mu4}  \label{1-2}
\end{equation}
and then defining Poincar\'{e} momenta by
\begin{equation}
p_\mu = \lim_{R\to\infty} \Pi_\mu.  \label{1-3}
\end{equation}
With these definitions the commutation relations of $l_{\mu4}$ are
\begin{equation}
[ \Pi_\mu, \Pi_\nu ] = \frac{-i}{R^2} l_{\mu\nu}  \label{1-4}
\end{equation}and, therefore,
\begin{equation}
[ p_\mu, p_\nu ] = 0.   \label{1-5}
\end{equation}

F. G\"ursey \cite{fg} has formulated the analogue of eq.\ (\ref{1-2}) 
for spin-1/2 representations
\begin{equation}
\Pi_\mu = \frac{1}{R} l_{\mu4} + \frac{1}{2R} \gamma_\mu   \label{1-6}
\end{equation}
where $\gamma_\mu$ are the usual Dirac matrices. 
(See also C. Fr{\o}nsdal et al\cite{ff}.)
The $\gamma$-term was named ``momentum spin" by F. G\"ursey.
If we measure the momentum in units of $1/R$ we can drop this
factor and rewrite (\ref{1-6}) in the form 
\begin{equation}
\Pi_\mu =  l_{\mu4} +  \frac{1}{2}\gamma_\mu .  \label{1-7}
\end{equation}

For $R\to\infty$ $l_{\mu4}$ will grow proportionally to $R$ and converge to 
$p_\mu$. $\gamma_\mu$ then becomes a second order term compared to $l_{\mu4}$.
In a more realistic application of anti de~Sitter space as a physical model of 
spacetime, $R$ is large but nevertheless finite. 
In this situation $p_\mu$ can be used as an approximation to $\Pi_\mu$ with 
$\gamma_\mu$ as a second order correction to $p_\mu$.  

Assuming a large but finite $R$, the following analysis will be based on an 
approximative translation operator of the form
\begin{equation}
t_\mu =  p_\mu +  \frac{1}{2}\gamma_\mu.   \label{1-8}
\end{equation}
We will deliberately ignore any other higher order contributions and 
determine the effect of the second order term $\gamma_\mu$ within the 
kinematics of a spin-1/2 particle system.

In the preceding lines, we have used the term of spacetime in order to 
establish the connection to the ``historical" application of de~Sitter 
symmetry. This does not imply that we will make any assumptions about the 
symmetry of spacetime in the large. In the following de~Sitter symmetry is 
assumed only as a symmetry of (angular) momentum space.

\section{AN INVARIANT OPERATOR OF A MULTIPARTICLE SYSTEM}

We start from a system of spin-1/2 particles in a tangential spacetime 
with a Minkowski structure. 
The multiparticle state space is given by the direct product of one-particle 
state spaces, which in turn are defined by momentum eigenstates that satisfy
the Dirac equation.  
We will call these particles for short Dirac particles and thereby mean
massive, structure less, lepton-like particles.

It is obvious that we can describe our multiparticle state space without 
loss of generality equally well by the direct product of two-particle states 
that belong to irreducible representations of P(3,1) (assuming an even number 
of particles). 

A well-known ``invariant" operator of P(3,1) is the modulus  $p_{\mu} p^{\mu}$
of the translation operator. It is invariant under all operations of P(3,1) 
and can, therefore, be represented by a fixed c-number within a given 
irreducible representation of P(3,1). 
A well-known consequence of this fact is, for example the Klein-Gordon 
equation. 
This property is not restricted to single particles but is also true for any 
isolated system of particles.

Consider now the operator $j_{ab} j^{ab}$, $a,b = 0,1,2,3,4$ (summation over
$a,b$), where $j_{ab}$ are spin-1/2 representations of the infinitesimal 
generators of SO(3,2). 
For an isolated system, which is described by an irreducible representation 
of SO(3,2), application of this operator again delivers a fixed c-number.
We evaluate this operator for a multiparticle system by using (\ref{1-8}) 
as an approximation for $j_{\mu4}$ and taking (\ref{1-1}) into account, 
collecting terms of the magnitude $R^2$ and $R$ and ignoring terms of lower 
magnitude. 
Our multiparticle state space in the tangential spacetime will 
serve as a 0th-order approximation in a perturbation expansion. 
We then obtain the expression
\begin{eqnarray}
& & p_\mu p^\mu + 2 p_\mu p'^\mu +  p'_\mu p'^\mu \nonumber \\
& & + \frac{1}{2}(\gamma_\mu p^\mu + \gamma_\mu p'^\mu + \gamma'_\mu 
p^\mu + \gamma'_\mu p'^\mu) \nonumber \\
& & + \cdots , 				\label{2-1}
\end{eqnarray}
which is identical to the modulus of the total translation operator
\begin{equation}
T_\mu = t_\mu + t'_\mu + \cdots    	\label{2-2}
\end{equation}
in this approximation that we are considering.

For an irreducible system this operator must deliver a fixed c-number. 
In other words, the modulus of the total translation operator is a 
constant-of-motion: 
\begin{equation}
T_{\mu} T^{\mu} = \mbox{const.} 	\label{2-3}
\end{equation}
Constants-of-motion are a useful means to study the internal kinematics of 
a physical system. So let us see.

Since the translation operator commutes with the total momentum $P_\mu$, 
its modulus $P^\mu P_\mu$ is also a constant. 
This enables us to separate the momentum terms that are quadratic in $p_\mu$
from (\ref{2-1}). 
(Later we will find a more explicit justification for this step.) 
Therefore, we obtain another constant expression
\begin{eqnarray}
& & \gamma_\mu  (p^\mu  + p'^\mu + \cdots)  \nonumber \\ 	\label{2-4}
&+& \gamma'_\mu (p'^\mu + p^\mu  + \cdots)  \nonumber \\
&+& \cdots 				    \nonumber \\ 				
&=& \mbox{const.} . 			
\end{eqnarray}

Here we have terms that represent the operators of a Dirac equation of 
individual particles and other terms - like $\gamma_\mu p'^\mu$ - 
that provide for a connection between pairs of particles. 
Whereas the former belong to a Poincar\'{e} covariant
description of a multiparticle system in Minkowski spacetime, the latter
can be understood as a perturbation to the Poincar\'{e} covariant system.
So we rearrange the terms
\begin{eqnarray}
& &  \gamma_\mu  (p^\mu  + a^\mu) \nonumber \\			\label{2-5}
&+&  \mbox{similar terms for the other particles} \nonumber \\
&=&  \mbox{const.} 
\end{eqnarray}
with the perturbation term
\begin{equation}
a^\mu = \sum p'^\mu  \label{2-6}
\end{equation}
(summation over all particles except the first).
We will analyze the effect of the term $a^\mu$ within the framework of a 
perturbation approach.

\section{STRUCTURE OF TWO-PARTICLE STATES }

Since the perturbation term $\gamma_\mu a^\mu$ is basically a two-particle 
operator, we will have to make use of two-particle states. So let us spend 
a short look at its general structure (ignoring spin variables).

Let 
\begin{equation}
|{\mathbf{P}}\rangle = \int \frac{d^3p}{p_0}\, \frac{d^3p'}{p'_0}\, 
C(\mathbf{p,p'})\, |\mathbf{ p, p'} \rangle   \label{3-1}
\end{equation} 
be a two-particle state with momentum $P$ of a two-particle representation 
of P(3,1) with $P^2 = M^2$ in a state space ${\mathcal{H}}_M$. 
The two-particle states $|\mathbf{p, p'}\rangle$ belong to the direct 
product of one-particle states $|\mathbf{p}\rangle$. 

With $p = k - q$, $p' = k + q$, $2k = P$ we can rewrite
\begin{equation}
|{\mathbf{P}}\rangle = \int \frac{d^3q}{q_0}\, \, \tilde{C}(\mathbf{q}) \,  
 |\mathbf{k - q, k + q} \rangle . 	\label{eq32}
\end{equation}
From
\begin{equation}
p^2 = p'^2 = m^2,  		\label{3-2}
\end{equation} 
where $m$ is the particle mass, and
\begin{equation}
P^2 = (p + p')^2 = M^2   \label{3-3}
\end{equation} 
we obtain
\begin{equation}
q^2 = m^2 - \frac{1}{4}M^2,  		\label{3-4}
\end{equation} 
\begin{equation}
q_0^2 = m^2 - \frac{1}{4}M^2 + {\mathbf{q}}^2  \label{3-5}
\end{equation} 
and 
\begin{equation}
kq = 0 . 			\label{3-6}
\end{equation} 

Conditions (\ref{3-4}) to (\ref{3-6}) express the fact that 
$|{\mathbf{P}}\rangle$ is a state of an two-particle representation 
characterized by the total mass $M$.

We can formulate the state (\ref{3-1}) also in terms of wave functions
\begin{equation}
e^{-iPx} = \int \frac{d^3q}{q_0}\, \, \tilde{C}({\mathbf{q}}) 
\,\, e^{-i(k-q)x} \, e^{-i(k+q)x} ,  \label{3-7}
\end{equation}
which are obtained by formally multiplying the ket-states 
$|{\mathbf{P}}\rangle$, $|{\mathbf{k-q}}\rangle$ and $|{\mathbf{k+q}}\rangle$
by its associated bra-states $\langle x|$.

The momenta $p$ and $p'$ of each term under the integral in (\ref{3-1}) 
adds up to $P$. 
This ``entanglement" of momenta within two-particle states in 
${\mathcal{H}}_M$ will be essential for the results that we will obtain.

\section{FORMULATION IN FOCK SPACE }

Let us now return to a multiparticle system. We formulate it with the help 
of standard Fock space methods. 
The ``free" part of our system is easily converted into a Fock space 
formulation following the usual ``quantization" of the Dirac field 
(see \cite{gs} or any text book on quantum field theory). 
The field operator of the Dirac field (taken from this reference) has
the form
\begin{equation}
\psi(x) = (2\pi)^{-3/2} \!\!\!\int\!\! d^3p 
\left( b_s({\mathbf{p}}) u_s ({\mathbf{p}}) e^{-ipx} 
\!+\! {d_s({\mathbf{p}})}^\dagger v_s({\mathbf{p}}) 
e^{ipx} \right). \label{4-1}
\end{equation}
A similar expression defines the Dirac adjoint operator $\bar{\psi}(x)$. 
$b^\dagger_s({\mathbf{p}}), b_s({\mathbf{p}})$ are electron emission 
and absorption operators, $d^\dagger_s({\mathbf{p}}), d_s({\mathbf{p}})$ 
are the corresponding operators for positrons.
They satisfy the usual anticommutation relations of the Dirac field.

As a first attempt we represent our two-particle perturbation terms 
$\gamma_\mu p'^\mu$ in Fock space in the following form  
\begin{equation}
\int d^3x\,d^3x' \, \bar{\psi}(x)\gamma_\mu \psi(x) \,\, 
\bar{\psi}(x') p^\mu \psi(x') . \label{4-2}
\end{equation}
This Fock space operator is not yet adapted to its immediate insertion into a 
perturbation calculation.
Remember that a quantum mechanical perturbation calculation requires the
combination of a perturbation term with a projection operator onto the basic 
state space. Therefore, we still have to incorporate a suitable projection 
mechanism onto our two-particle basic.

We will achieve this by collecting only those terms of (\ref{4-2}) that 
contribute if we evaluate this operator for two-particle states with momentum 
$P$ of a given (irreducible) two-particle state space ${\mathcal{H}}_M$.

Consider the following contribution to (\ref{4-2})
\begin{equation}
\dots {\bar{b}({\mathbf{p+q}}) \gamma_\mu} {b({\mathbf{p}})\,\,} 
{\bar{b} ({\mathbf{p'-q'}}) p^\mu} b(\mathbf{p'})\dots . 
\label{4-3}
\end{equation}
If we evaluate this operator for a two-particle state, then only such terms
with $\mathbf{q=q'}$ will be involved, as a consequence of momentum 
entanglement within two-particle states.
 
This is true for every state of ${\mathcal{H}}_M$ with a given momentum $P$. 
And since every two-particle state of ${\mathcal{H}}_M$ can be represented by 
a superposition of two-particle momentum eigenstates, it is generally valid.
So we can drop the restriction to a fixed $P$ and collect all contributions 
that belong to the same $\mathbf{p}$ and $\mathbf{q}$. 
Hence, we can write 
\begin{equation}
\dots {\bar{b}({\mathbf{p + q}}) \gamma_\mu b ({\mathbf{p}})} \, 
\tilde{a}^\mu({\mathbf{q}}) \dots  \label{4-4}
\end{equation}
with
\begin{equation}
\tilde{a}^\mu({\mathbf{q}}) = \int dV(p')
\,\bar{b}({\mathbf{p' - q}}) p^\mu b ({\mathbf{p'}}), \label{4-5}
\end{equation}
where $dV(p')$ indicates a summation over all contributions that belong to
the same $\mathbf{p}$ and $\mathbf{q}$. 
An analogous consideration is valid for positron and mixed terms.

Let us analyze the meaning of $\tilde{a}^\mu(\mathbf{q})$ in more detail. 
If we evaluate (\ref{4-4}) for a two-particle state, the second particle 
will contribute a complex amplitude given by the expectation value of 
$\tilde{a}^\mu(\mathbf{q})$ that acts as a multiplicative weight to the 
expectation value of the first particle term. 
This weight depends on $\mathbf{q}$ and fully describes the contribution of 
a second particle to the total expectation value. A long as our focus is on 
the first particle, then all we need to know about other particles are the 
complex weights that apply to the expectation values of particle one. 
This fact simplifies the treatment of a single particle within an ambient 
multiparticle system. 
To keep track of the weighting factors that apply to each ``transition" 
$p \rightarrow p+q$ in (\ref{4-4}) we need a suitable ``bookkeeping" system.

We can establish such a bookkeeping system by introducing an auxiliary Fock 
space with operators that emit and absorb quanta with momentum $\mathbf{q}$. 
If we prepare a state in this Fock space by applying an emission operator
multiplied by a complex amplitude onto the vacuum state, then a later
application of an absorption operator will redeliver this amplitude. 
This is exactly what we need. 

We have some freedom in doing this, as long as the system is able to keep 
track of the amplitudes of the momenta $\mathbf{q}$.
So let us replace the operator $\tilde{a}^\mu(\mathbf{q})$ of (\ref{4-5}) 
by a new operator $a^\mu(\mathbf{q})$ of our bookkeeping system 
\begin{equation}
\tilde{a}^\mu({\mathbf{q}}) \rightarrow e \, a^\mu ({\mathbf{q}}). 
\label{4-6}
\end{equation}
We have to insert a - so far unknown - normalization factor $e$, because
the commutation relations that we will introduce below define a normalization 
of $a^\mu$, whereas (\ref{4-5}) defines a different normalization of 
$\tilde{a}^\mu$.
The determination of $e$ in the next section will involve a detailed 
examination of the integration volume in (\ref{4-5}).

We define the following commutation relations between $a_\mu$ and its ajoint
operator $a_\mu^\dagger$
\begin{equation}
[a^\mu({\mathbf{k}}), a^\nu ({\mathbf{k'}})^\dagger] 
= \delta({\mathbf{k - k'}}) . \label{4-7}
\end{equation}
Then $a^{\mu\dagger}(\mathbf{k})$ are emission operators and 
$a^\mu(\mathbf{k})$ absorption operators for quanta with momentum 
$\mathbf{k}$. 

Following the usual procedure in the ``quantization of the radiation field"
(see e.g. \cite{gs}) we define the operators
\begin{eqnarray}
A^j(x) = (2\pi)^{-3/2} \!\!\!\int\!\! \frac{d^3k}{k^0 \sqrt{2}} 
\left( a^j({\mathbf{k}}) e^{-ikx} 
+ a^j({\mathbf{k}})^\dagger e^{ikx} \right), \nonumber \\
j=1,2,3, \label{4-8}
\end{eqnarray}
and
\begin{equation}
A^0(x) = (2\pi)^{-3/2} \!\!\!\int\!\! \frac{d^3k}{k^0 \sqrt{2}} 
i \left( a^0({\mathbf{k}}) e^{-ikx} 
+ a^0({\mathbf{k}})^\dagger e^{ikx} \right). \nonumber \\
\label{4-9}
\end{equation}
$k^0$ shall be determined by condition (\ref{3-5}) if these operators are 
evaluated within two-particle states. 
(In the "free radiation field" $k^0$ is ``on-shell": $k^0 = |\mathbf{k}|$.)  

If we add the spacetime dependencies to the emission and absorption operators
in (\ref{4-4}), as prescribed by (\ref{3-7}), we obtain
\begin{equation}
\dots \bar{b}({\mathbf{p+q}}) e^{i(p+q)x} \gamma_\mu 
b({\mathbf{p}}) e^{-ipx}\,\, 
a^\mu({\mathbf{q}}) e^{-iqx} \dots . \label{4-10}
\end{equation}
Note that the correct spacetime dependency of $a^\mu$ is obtained from 
(\ref{4-5}).

After inserting the spin functions $u_s(\mathbf{p})$ and $v_s(\mathbf{p})$
these terms and the corresponding positron and mixed terms add up to a 
Fock operator in the form
\begin{equation}
e \int d^3x\, : \bar{\psi}(x)\gamma_\mu \psi(x) : A^\mu(x), \label{4-11}
\end{equation}
where $::$ stand for normal ordering of emission and absorption operators.  

The bookkeeping system has enabled us to give our perturbation term 
the same structure as the interaction term of quantum electrodynamics 
(QED) with the constant $e$ acting as a coupling constant. 
Therefore, the full apparatus of QED is available to analyze the 
effect of this term in a perturbation calculation. The result of such
an analysis is well-known and strongly suggests that a Dirac particle 
within our SO(3,2) model shows properties of an electrically charged 
particle. 
 
Let us come back to the separation of the quadratic terms in (\ref{2-1}) 
from those that are linear in $p$: 
Our perturbation term provides the exchange of momenta between two particles 
without changing the total momentum. Therefore, also the square of the
total momentum is conserved. This justifies the separation. It also adds to 
our understanding why it is possible to observe small terms (linear in $R$) 
in the presence of large terms (quadratic in $R$).

The standard formulation of QED introduces the interaction term by the
postulation of gauge invariance. The latter is achieved by adding a new
"gauge field" that couples to the electron field. Despite the great success 
of the gauge principle it does not really provide us with an insight into
the interaction process itself.
We, on the other hand, have explicitly constructed an interaction term from 
known elements of the electrons Fock space and also have defined the 
bookkeeping field in terms of these elements. This gives us a full
understanding of the mathematical and physical nature of this interaction
term.

Unlike the standard formulation, where the coupling constant enters as 
a free parameter that has to be determined by the experiment, our approach
does not leave room for any free parameter. This means that the coupling 
constant $e$ is determined by the theory and, therefore, should be calculable.

\section{ESTIMATE OF THE COUPLING CONSTANT}

This coupling constant is defined by the normalization of 
$\tilde{a}^\mu({\mathbf{q}})$ in (\ref{4-5}) relative to the  
operator $a^\mu({\mathbf{q}})$ whose normalization is given by (\ref{4-7}).
This ratio can - in principle - be determined by correctly ``counting" all 
contributions to (\ref{4-5}) - in other words: by a careful analysis of the
volume element of the integral in (\ref{4-5}). 

More than 30 years ago A. Wyler \cite{aw} discovered that the 
fine-structure constant $\alpha$ can be expressed by volumes of certain 
symmetric spaces. Being a mathematician he was not able to put his 
observation into a convincing physical context. 
Therefore, his work was criticized as fruitless numerology \cite{br}.  

Wyler's idea was picked up later by F. D. Smith, Jr. \cite{fds} who extended 
Wyler's heuristic approach into a general scheme based on a 
fundamental Spin(8) symmetry. 
Smith was then able to express coupling constants and relations of particle 
masses by characteristic volumes with a remarkable degree of precision. 

Let us see how far our model will lead us and whether we possibly can find a 
physical explanation for these authors' observations. 

Consider the S-matrix element of electron-electron scattering (M{\o}ller 
scattering) (see e.g. \cite{gs}). In the conventional formulation of QED this 
term involves two vertices and two $a^\mu$-operators. 
The latter result in an intermediate photon. It defines those intermediate 
states that contribute to the calculation of the scattering amplitude.
In our formulation we have two $\tilde{a}^\mu$-operators given by the
integral (\ref{4-5}) instead. Whereas, by definition, there is one and only 
one $a^\mu(\mathbf{q})$ for each $\mathbf{q}$, the multiplicity of 
$\tilde{a}^\mu(\mathbf{q})$ with respect to $\mathbf{q}$ may be different.
Our task will be to determine the multiplicity of $\tilde{a}^\mu(\mathbf{q})$ 
relative to $a^\mu(\mathbf{q})$ by ``counting" the intermediate states that 
our approach allows for. 

We already have parameterized the contributions to the interaction term by 
the parameters $q$ and $p'$. 
The way in which the parameters $q$ are used in the perturbation calculation
defines the parameter space of $q$ as (a subspace of) the euclidean $R^3$. 
If we split off the $q$-dependency we are left with the integral over $p'$
and our task will be to determine the multiplicity, or the integration volume 
respectively, of the contributions with respect to $p'$. 

The basis for the evaluation of the integration volume is the particle 
momenta and the homogeneous Lorentz group acting on the particle momenta. 
The SO(3,1) acts transitively on a particles mass shell
\begin{equation}
p_0^2 -p_1^2 - p_2^2 - p_3^2 = m^2 . \label{5-1}
\end{equation} 
The independent parameters $p_1,p_2,p_3$ span a 3-dimensional parameter space. 
For a two-particle state of a representation with mass $M$ we have instead
\begin{equation}
(p_0+p_0')^2-(p_1+p_1')^2-(p_2+p_2')^2-(p_3+p_3')^2 = M^2. \label{5-2}
\end{equation}
We can convert this into
\begin{eqnarray}
p_0^{2}+p_0^{'2}-p_1^{2}-p_1^{'2}-p_2^{2}
-p_2^{'2}-p_3^{2}-p_3^{'2}          \nonumber \\
+2p_0 p_0'-2p_1 p_1'-2p_2 p_2'-2p_3 p_3' = M^2.	\label{5-3}
\end{eqnarray}
From (\ref{5-1}) and (\ref{5-2}) follows that
\begin{equation}
p_0 p_0'-2p_1 p_1'-2p_2 p_2'-2p_3 p_3' = \kappa^2   \label{5-4}
\end{equation}
must be invariant.
Therefore,
\begin{eqnarray}
p_0^{2}+p_0^{'2}-p_1^{2}-p_1^{'2}-p_2^{2}-p_2^{'2}-p_3^{2}-p_3^{' 2} 
\nonumber \\
 = M^2 - \kappa^2 .	\label{5-5}
\end{eqnarray}
The symmetry group of this quadratic form is SO(6,2). 
Relation (\ref{5-4}) reduces the number of independent parameters from 6 to 5 
and thereby SO(6,2) to SO(5,2). 
SO(5,2) acts transitively on this 5-dimensional parameter space.
Each point in this parameter space corresponds to a state in the two-particle 
state space ${\mathcal{H}}_M$. Therefore, the volume of the parameter space
delivers a measure for the number of states that can contribute to the
interaction term.

Given a point $Q$ in this parameter space, then other points can be reached 
by applying a linear transformation of SO(5,2) to $Q$. There are certain 
transformations that do \begin{em}not\end{em} change the point $Q$. 
These transformations form the subgroup S(O(5) x O(2)).
This is the isotropy subgroup or stabilizer of $Q$. Therefore, to obtain the
multiplicity of states, we have to start from the coset space 
$D_5$ = SO(5,2)/S(O(5) x O(2)) rather than from SO(5,2).  

$D_5$ is a symmetric space. 
By construction $D_5$ is isomorphic to ${\mathcal{H}}_M$.
It is known from the work of Hua and Lu \cite{hl}
that $D_5$ can be represented by matrices;
that is, this symmetric space is isomorphic to the real hyperball
\begin{equation}
{\mathcal{R}}_R(5,2) = \{X\in R^{5\times2} \,|\, I-XX' > 0 \}. \label{5-6}
\end{equation}
(See \cite{fkklr} for a modern introduction to symmetric spaces.)
Hua \cite{lkh} has calculated volumes of ${\mathcal{R}}_R(5,2)$ and other
domains.
In contrast to ${\mathcal{R}}_R(5,2)$ $D_5$ has an infinite volume.

Consider now a two-particle state $|{\mathbf{P}}\rangle$ with momentum $P$. 
Then there is another volume associated with $D_5$. 
This is the subspace of all points that correspond to 
a situation where for one of the particles $p_0 = m$ and the other particle 
has reached its maximum value of $p'_0 = P_0 - m$. 
For reasons of symmetry this volume is spherical symmetric and isomorphic to 
the border sphere $C_5$ of $D_5$. $C_5$ has 4 dimensions.
Then all states with given $P_0$ are confined to a volume $\bar{D}_5$ inside 
of $C_5$ and including $C_5$.
The subspace $\bar{D}_5$ of $D_5$ is finite and can be mapped onto 
${\mathcal{R}}_R(5,2)$ by an isometric mapping.

If $Q = (q_1,...,q_5)$ is a point of $\bar{D}_5$ that is mapped into
a point $S = (s_1,...,s_5)$ of ${\mathcal{R}}_R(5,2)$, then we can 
establish a one-to-one relationship such that 
\begin{equation}
q_i = r \, s_i,    \label{5-7}
\end{equation}
where $r$ is a properly chosen scaling factor. This gives us the choice
to use either $q_i$ in $\bar{D}_5$ or $s_i$ in ${\mathcal{R}}_R(5,2)$
as integration parameters. 

To be consistent with Smith's terminology we will calculate all volumes in 
${\mathcal{R}}_R(5,2)$. We will use the notation $V(D_5)$ for the volume 
that corresponds to $\bar{D}_5$ but is calculated in ${\mathcal{R}}_R(5,2)$ 
and will remember that we have to apply the correct number of scaling 
factors $r$.

$C_5$ has another important property: If particle 1 is initially at rest
and a second particle with a given momentum $p'$ is added to form a
two-particle state, then this state corresponds to a point on $C_5$ as 
described before. Other states can be generated from this ``initial" state
by the exchange of momentum. Therefore, to determine all states that 
are eventually involved we have first to collate all initial states. 
This means, we have to perform an integration over $C_5$ with a volume 
element $d^4s/V(C_5)$.
This delivers a first volume factor of $1/V(C_5)$.

To collect all possible momentum changes of particle 1 we have to integrate
over $\bar{D}_5$. From condition (\ref{3-6}) it follows that for a given 
$|{\mathbf{P}}\rangle$ only momentum exchanges in the subspace perpendicular 
to $P$ have to be considered. 
Since the direction of the total momentum is undetermined (when we are 
constructing the interaction operator), we have to keep the integration 
over $D_5$.  
We compensate for this by a volume factor of $1/V(S_4)$ where 
$S_4 = SO(5,2) / SO(4,2) = SO(5) / SO(4)$ is the unit sphere in 4 dimensions. 
This reduces the number of independent parameters to 4. 
Let $(s_1,..,s_4)$ be a new set of independent parameters corresponding to
a new set $(q_1,..,q_4)$.  

If we integrate over $\bar{D}_5$ using this new parameter set, each $s_i$ 
will be responsible for a contribution of $V(D_5)^{1/4}$ to the volume of 
$\bar{D}_5$ .
Three of these parameters can now be mapped onto the transferred momentum $q$.
The fourth parameter $s_4$, obviously, corresponds to a momentum transfer 
within each of the particle momenta, without any momentum transfer between 
the particles. Such transitions contribute to the volume of $C_5$.
We can perform the integration over $s_4$ and obtain a correcting factor 
to the already calculated volume $V(C_5)$ of $V(D_5)^{1/4}$.

There are two more factors that contribute to the multiplicity of momentum
states. One is related to the spin components of the particle states, which
give each momentum state a multiplicity of $4\pi$ because of the periodicity
of spin states. The other factor is related to the (relative) phases of the 
momentum states within multiparticle states. By adding another factor of 
$2\pi$ we take into account this degree-of-freedom.

After extracting these constant factors from the integral we are left with an 
integration over the $p'$ parameter space where now the integrand should enter 
with a multiplicity of one within the $p'$-parameter space - provided that we 
have correctly captured all factors that determine any multiplicities. 
Collecting these factors we end up with  
\begin{equation}
8 \pi^2 \,V(D_5)^{1/4} \, / \, (V(S_4) \, V(C_5)).    	\label{5-8}
\end{equation}
This is essentially Wyler's formula.

The application of (\ref{4-5}) to a state $|\mathbf{p}\rangle$ now describes a 
transition to $|\mathbf{p-q}\rangle$ with a weight given by the square root of 
(\ref{5-8}). (Remember that in our estimation (\ref{4-5}) has entered twice.) 
This demonstrates that the weight factor has the property of a coupling 
constant.

The volumes $V(D_5)$ and $V(C_5)$ have been calculated by 
L. K. Hua \cite{lkh}. 
$V(S_4)$ is the volume of the unit sphere $S_4$ in 4 dimensions. With
\begin{equation}
V(C_5) = \frac{8 \pi^3}{3},        			\label{5-9}
\end{equation}
\begin{equation}
V(D_5) = \frac{\pi^5}{2^4\, 5!},   			\label{5-10}
\end{equation}
\begin{equation}
V(S_4) = \frac{8 \pi^2}{3}         			\label{5-11}
\end{equation}
we obtain 
\begin{equation}
\frac{9}{8 \pi^3} \left(\frac{\pi^5}{2^4 \, 5!}\right)^{1/4} . 	\label{5-12}
\end{equation}
If we identify this value with the coupling constant 
$e^2/(2\pi)^2 = \alpha/\pi$ in the S-matrix element for M{\o}ller scattering 
we obtain a value for 
$\alpha$
\begin{equation}
\alpha = 1/137.03608245.   				\label{5-13}
\end{equation}
Although intended only as an estimate, this result is in agreement with 
experimental values like 1 : 1.0000005. 
(A value of 137.035 999 93(52) has been determined from the magnetic moment 
of the electron \cite{kino}.)

We can easily convince ourselves that the scaling factors $r$ either cancel 
or are absorbed in the volume element $d^3q$.

\section{CONCLUSION }

A perturbational analysis of G\"ursey's momentum spin has led us to a 
mathematical structure that is identical to the perturbation theoretical 
formulation of quantum electrodynamics. 
We have found an estimate of the coupling constant that is in excellent 
agreement with the experimental values of the electromagnetic coupling 
constant.

It is, therefore, very tempting to identify G\"ursey's momentum spin with 
the current of quantum electrodynamics. This opens a new and simple access 
to the electromagnetic interaction in terms of standard quantum mechanics of 
multiparticle systems:

\begin{em}Electromagnetic interaction\end{em}: Effects of a quantum 
mechanical operator term that takes care of the momentum spin in a 
perturbation treatment. It connects states of different momenta and,
thus, causes transitions between these states.

\begin{em}Coupling constant\end{em}: A constant that results from counting 
the possible intermediate states if we evaluate the interaction operator for 
two-particle states.
It delivers a kind of mathematical signature for our construction of the 
interaction term. 

\begin{em}Photon\end{em}: Not a real physical particle but a mathematical 
bookkeeping item that keeps track of exchanged momenta, with ``off-shell" 
behaviour.

\begin{em}Radiation field\end{em}: Those photons that are exchanged over long 
distances. For such distances, the formalism of QED tells us that the  
exchanged momentum will be placed on the mass zero shell.

\begin{em}Off-shell particles\end{em}: Off-shell behaviour of Dirac particles 
shows up very clearly as a property of the anti-commutation functions, rather 
then of particle states. 

\begin{em}Quantized fields\end{em}: There is no quality that goes beyond the 
properties of plain Fock space operators that are used to describe the 
multiparticle system of electrons and to build up a bookkeeping system for 
the exchanged momenta. 

\begin{em}Vacuum polarization\end{em}: A pictorial description of certain
Feynman graphs without any real physical background. There is no ``vacuum" 
other than the bare vacuum state of the Fock space representation.

\begin{em}Pair creation\end{em}: Not really a creation process, but rather 
a transition from an anti-particle state (moving backward in time) to a 
particle state (moving forward in time).

\begin{em}Locality\end{em}: Locality in the sense of interaction at a single 
point: the construction process of the interaction operator shows us how 
this follows directly from the structure of two-particle states which 
we have used as a basis.

\begin{em}Causality\end{em}: Although our SO(3,2) model can be understood
as an action-at-a-distance theory we know very well that the formalism of 
QED is in agreement with relativistic causality. 

\begin{em}Gauge invariance\end{em}: Gauge invariance appears as a 
consequence of the existence of a charged current rather than vice versa. 

\begin{em}Poincar\'{e} invariance\end{em}: An approximation to a basic
SO(3,2) symmetry that is observed when the contributions of momentum spin 
compensate in systems with an equal amount of positive and negative electrical 
charges. Because of this compensation the Poincar\'{e} invariance is usually
realized better than the translation operator (\ref{1-8}) suggests.

We have been led to a theory of interacting fields by a ``slight" 
modification of the basic symmetry: we have replaced the Poincar\'{e} group 
P(3,1) by a de~Sitter group SO(3,2) and used the Poincar\'{e} symmetry as an
approximation to the de~Sitter symmetry. 
No additional assumptions - except basic principles of quantum mechanics - 
have been employed. 

In contrast to the standard model, the coupling constant is uniquely 
determined by the theory. 

We have deliberately ignored contributions of other higher order terms. 
Such terms still deserve more attention. They may lead to the formulation
of different types of interaction.

\end{document}